\documentstyle[12pt,aaspp4]{article}
\begin{document}
\parindent=1.0cm

\title{The Central Regions of M81}

\author{T. J. Davidge \& S. Courteau \altaffilmark{1} $^{,}$ \altaffilmark{2}}

\affil{Herzberg Institute of Astrophysics, \\ National Research Council of 
Canada,\\ 5071 W. Saanich Road, Victoria, BC Canada V8X 4M6\\ {\it email: 
tim.davidge@hia.nrc.ca, stephane.courteau@hia.nrc.ca}}

\altaffiltext{1}{Visiting Astronomers, Canada-France-Hawaii telescope, which is 
operated by the National Research Council of Canada, the Centre National de la 
Recherche Scientifique, and the University of Hawaii}

\altaffiltext{2}{Guest Users, Canadian Astronomy Data Centre, which is 
operated by the Herzberg Institute of Astrophysics, National Research Council 
of Canada}

\begin{abstract}

	High angular resolution near-infrared images, obtained with the 
Canada-France-Hawaii Telescope adaptive optics system, are combined with 
archival Hubble Space Telescope data to investigate the central regions of 
the nearby Sb galaxy M81 (NGC3031). The spectral-energy distribution of the 
circumnuclear region, which extends out to $1\farcs5$ 
($\sim 24$ pc if $\mu_0 = 27.5$) from the nucleus, can be modelled 
as a combination of an old metal-rich population and 
emission from hot dust. Thermal emission has been detected 
near other AGN, and simple models indicate that hot dust can account 
for $\sim 20\%$ of the light in $K$ within $0\farcs5$ of the 
M81 nucleus. An elongated structure with M$_V \sim -7$, which may 
be an area of active star formation, is detected 0$\farcs$45 from the nucleus. 
At distances in excess of 1$\farcs$5 from the nucleus the $J-K$ color of the 
M81 bulge is not significantly different from what is seen in M31. The HST data 
are also used to search for bright globular clusters within 2kpc of 
the center of M81. The area within 0.26 kpc of the M81 nucleus is 
largely devoid of bright globular clusters, in agreement with what is seen 
in the central regions of the Galaxy and M31. However, our survey indicates 
that there may be $\sim 45 \pm 12$ globular cluster candidates with M$_V \leq 
-7$ within 2 kpc of the galaxy center, which is consistent with 
what would be infered from the Milky-Way cluster system 
after adjusting for differences in the total number of clusters.

\vspace{0.2cm}
\noindent{\it Key Words: galaxies: individual (NGC3031) -- galaxies: nuclei -- 
galaxies: Seyfert -- galaxies: star clusters}

\end{abstract}

\section{INTRODUCTION}

	The central region of a galaxy lies at the bottom of a deep 
gravitational well, and physical processes such as dynamical friction, 
accretion, and dissipation of angular momentum can cause material to accumulate 
in this area that may provide insight into the past history of the host galaxy. 
In fact, there are indications that the evolutionary processes that defined the 
central characteristics of galaxies were major events that had a profound 
influence on the host systems (e.g. Faber et al. 1997). However, efforts to 
survey the stellar content near galaxy centers are frustrated by extremely 
high stellar densities, which not only make the detection of individual 
objects difficult, but can create environments that are systematically 
different from the solar neighborhood, thus complicating the interpretation of 
the results, even in the nearest systems. For example, the strength of 
$2.3\mu$m CO absorption weakens within $\sim 10$ arcsec of SgrA* (Haller 
{\it et al.} 1996, Sellgren {\it et al.} 1990). There are a number of possible 
causes of this effect (Sellgren {\it et al.} 1990), including the 
stripping of stellar envelopes due to exposure to an intense radiation field, 
although current evidence favours the presence of an unresolved body of 
early-type stars (Eckart {\it et al.} 1995, Davidge {\it et al.} 1997a).

	The Galactic Center (GC) may not be representative of the 
nuclear regions in other spiral galaxies. The nearest external spiral galaxy, 
M31, shows a complicated nuclear morphology (Lauer {\it et al.} 1993) 
that differs from the GC (Davidge {\it et al.} 
1997b). While spectroscopic data are suggestive of an age gradient near the 
center of M31 (Davidge 1997), there is as yet no evidence for a population of 
very young stars, such as those in the SgrA complex, near the center of that 
galaxy. Clearly, it is important to survey the central regions of a larger 
sample of nearby galaxies at high angular resolutions to study nuclear 
morphology and search for evidence of recent star formation.

	The central regions of M81 are of interest because this galaxy 
is one of the closest objects outside the Local Group, and has a 
morphological type similar to M31 (Sandage \& Tammann 1987). In addition, the 
orientation of M81 leads to a clear view of the nucleus. Dynamical studies 
suggest that a central super-massive object is present (Keel 1989; 
Ho, Filippenko, \& Sargent 1996), and M81 is the closest galaxy to show the 
spectroscopic signatures of Seyfert-1 and LINER activity (Ho {\it et al.} 
1996; Keel 1989; Filippenko \& Sargent 1988). There is nuclear 
radio emission originating from a compact area having projected dimensions 
$\leq 700 \times 300$ AU (Bienlenholz {\it et al.} 1996), and 
Falcke (1996) and Reuter \& Lesch (1996) conclude that 
the M81 nucleus is a more energetic version of SgrA*. Spectroscopic studies 
predict that the narrow line region (NLR) is contained within a few arcsec 
(ie. a few tens of parsecs) of the nucleus 
(Filippenko \& Sargent 1988), and the asymmetric profiles of forbidden lines 
indicate that sub-structures should be present in this area (e.g. Keel 1989). 
However, previous investigations with the Hubble Space Telescope (HST) FOC and 
WFPC2 images at ultraviolet and visible wavelengths have shown only the central 
nucleus, with no evidence of sub-structure, star clusters, or a population of 
hot young stars (Devereux, Ford, \& Jacoby 1997).

	In the current paper, new high angular resolution near-infrared 
images are combined with existing HST data to explore 
the central regions of M81. Images recorded in the infrared 
are of interest since (1) they are relatively unaffected by all but the 
heaviest dust obscuration, (2) the contrast between 
the nuclear continuum emission, which has a blue spectral-energy 
distribution (SED), and the surrounding bulge is lower than at shorter 
wavelengths, simplifying the search for structures close to the nucleus, (3) 
there are prominent spectroscopic features, such as the $2.3\mu$m CO bands, 
that are useful probes of stellar content and can be studied with narrow-band 
filters, and (4) emission from warm dust, which has been seen around other 
active galactic nuclei (e.g. Alonso-Herrero, Ward, \& Kotilainen 1996), can be 
detected at these wavelengths.

\section{OBSERVATIONS AND REDUCTIONS}

	Images were recorded through $J, H, K$, Br$\gamma$, CO, and 
$2.26\mu$m continuum filters during the nights of UT March 7 -- 8 1998 
using the Canada-France-Hawaii Telescope (CFHT) Adaptive Optics 
Bonnette (AOB) (Rigaut {\it et al.} 1998) and KIR imager. KIR contains a 
$1024 \times 1024$ Hg:Cd:Te array with 0$\farcs$034 pix$^{-1}$. 
The CFHT AOB uses objects on the sky as reference beacons, 
and the M81 nucleus is sufficiently bright at visible 
wavelengths that it was employed for this purpose. A complete 
observing sequence consisted of either 2 or 4 exposures per filter at 4 dither 
positions that defined a $0\farcs 5 \times 0\farcs 5$ grid, and additional 
details of the observations are listed in Table 1. The FWHM entries in the last 
column of this table were measured from the point-like central nucleus 
of M81 after removing the underlying bulge light profile (\S 3). A 
field 1$^\circ$ North of the M81 center was observed periodically to monitor 
background sky brightness. Standard stars from Casali \& Hawarden (1992) were 
also observed throughout the two night observing run. The photometric 
calibration of the $J, H,$ and $K$ M81 observations was 
verified by simulating the aperture measurements made by Willner 
{\it et al.} (1985), and we find excellent agreement with their published 
colors.

	The data were reduced using the procedures described by Davidge {\it et 
al.} (1997a). The PSF in the final images consists of a bright central peak, 
with a width approaching that of the telescope diffraction limit for the images 
recorded near $2\mu$m, surrounded by an Airy pattern and 
faint satellite structures, which have been previously noted in AOB 
images by Chapman, Walker, \& Morris (1998). We suspect 
that the latter are the consequence of using an extended object as the AO 
reference source, as images of other fields, where stars were used as reference 
beacons, did not show these features. The absence of bright, isolated point 
sources in the M81 field prevented the use of deconvolution techniques to 
suppress the PSF satellites. Fortunately, although the presence of satellite 
structures in the PSF confounds efforts to search for faint sources close to 
the nucleus, the amplitude and location of the PSF satellites are not 
wavelength-sensitive, with the result that they are removed when images are 
ratioed. Consequently, the AOB data can be used to study the near-infrared SED 
of the M81 circumnuclear region at sub-arcsec angular scales.

	Archival HST WFPC2 F547W and NIC2 F160W images of the center of M81 
were extracted from the CADC HST database to search for faint sources close to 
the nucleus. The WFPC2 PC detector has a pixel scale of 0$\farcs$046 
pix$^{-1}$, and samples a field comparable in size to that observed by KIR. 
The WF chips have $0\farcs$1 pix$^{-1}$, and sample distances in excess of 
2 arcmin from the nucleus. The WFPC2 data, which consist of frames 
U2JQ0301T, 2T, 3T, and 4T from program GO-5433, were pipeline processed using 
the CADC recalibration and median-combined. The NIC2 detector has a pixel scale 
of 0$\farcs$075 pix$^{-1}$, so that the angular sampling is roughly a 
factor of two coarser than that offered by KIR. The NICMOS observations, 
which consist of frames N3ZD0NYTQ, UQ, and VQ from program GO-7331, were 
pipeline processed and median combined. 
 
\section{RESULTS}

\subsection{The Circumnuclear Environment of M81}

	The $J-K$ color profile near the nucleus of M81 
is shown in the top panel of Figure 1. The angular measurements plotted in this 
figure can be converted to spatial distances using 
the adopted distance modulus of $\mu_0 = 27.5$, so that $1\farcs 0 = 15.3$ pc. 
The $K$ image was gaussian-smoothed to match the angular resolution of 
the $J$ data, and the plotted points are azimuthal averages within 0$\farcs$2 
annuli. The error bars show the systematic uncertainties introduced by 
sky subtraction.

	$J-K$ changes significantly within a few arcsec of the M81 nucleus, 
with the blue nuclear continuum producing a local minimum in the color profile 
at the galaxy center. The actual $J-K$ color of the central source is likely 
much smaller than that measured here, as seeing blurs our data and flattens 
color gradients. $J-K$ reaches a maximum when $r = 0\farcs4$, 
and at distances between 0$\farcs$5 and 1$\farcs$5 from the nucleus 
$J-K$ declines with radius, before levelling off at larger radii. 

	The $J-K$ color of M81 when $r \leq 1\farcs 5$ is very different 
from that in the corresponding region of M31. This is demonstrated 
in Figure 1, where M31 $J-K$ measurements from Mould {\it et al.} (1989), 
scaled along the horizontal axis to account for the greater distance to M81, 
are plotted as open squares. This comparison indicates that the 
circumnuclear region in M81 has a relatively red color when compared with M31.
However, at distances in excess of 1$\farcs$5 from the nucleus, the mean $J-K$ 
color of the M81 bulge agrees with that in M31, indicating a 
similarity in stellar contents. 

	The $(J-H, H-K)$ two-color diagram (TCD), plotted in the top panel of 
Figure 2, provides further evidence that the circumnuclear region of M81 has a 
SED that is very different from what is typically near the centers of bulges 
and early-type galaxies. The M81 data for points beyond 1$\farcs$5 
from the nucleus (open squares) overlap with the elliptical and S0 aperture 
measurements (crosses) made by Frogel {\it et al.} (1978), and fall slightly 
below the Baade's Window (hereafter BW) M giant sequence (solid line) as 
defined by Frogel \& Whitford (1987). The M81 data for points within $1\farcs5$ 
of the nucleus, shown as filled squares, depart significantly from the BW M 
giant relation, and define a trend that is very different from 
the Galactic reddening vector, which is also shown in this figure. 
In \S 4 we demonstrate that the near-infrared SED of the circumnuclear 
region can be modelled as the sum of a stellar and non-stellar component, 
with the latter presumably due to hot ($T_{eff} \sim 1000$ K) dust.

	The narrow-band CO, Br$\gamma$, and $2.26\mu$m continuum images 
sample the longest wavelengths in the AOB dataset, and hence
have higher Strehl ratios than the broad-band images. The modest difference in 
central wavelength between these three filters also means that any structure in 
the PSF is removed when any two images are ratioed. 
The radial distribution of the CO index, defined as 
-2.5 log(CO/$2.26\mu$m continuum) plus a zeropoint, 
is shown in the middle panel of Figure 1, where 
the plotted values are azimuthal averages in 0$\farcs$1 annuli. 
The CO index weakens with decreasing radius when $r \leq 0\farcs 5$. This is 
likely due to veiling of the CO features by non-photospheric emission (\S 4), 
as the near-infrared colors at these radii argue against a large population 
of early-type stars being present. The CO index has values appropriate for old, 
metal-rich populations when r $\geq 0\farcs6$, indicating that the light is 
predominantly photospheric at these radii.

	The $(CO, J-K)$ TCD is shown in the lower panel of Figure 2. The CO 
indices plotted in this figure were measured after the CO and continuum images 
were gaussian smoothed to match the angular resolution of the $J$ observations; 
therefore, the CO indices in Figure 2, especially those at small radii, 
differ from those in the middle panel of Figure 1, which were measured from 
unsmoothed images. When $r \geq 1\farcs 5$ the M81 data 
occupy the same region of the TCD as early-type galaxies, and 
have smaller CO indices than bright giants in BW. At smaller radii the 
M81 data depart from the area occupied by early-type galaxies, and do not 
follow the trend defined by the reddening vector. As with the $(J-H, H-K)$ 
TCD, this behaviour can be modelled as a combination of stellar and non-stellar 
components (\S 4).

	Case B recombination theory predicts that the 
ratio of Br$\gamma$ and H$\alpha$ fluxes is roughly 0.01 
(Giles 1977, Brocklehurst 1971), so that Br$\gamma$ emission can be weak 
and difficult to detect. While the moderately short integration 
times used here are probably not sufficient to detect circumnuclear Br$\gamma$ 
emission in M81, observations with the Br$\gamma$ 
filter still provide a means of probing the SED near $2\mu$m 
at high angular resolutions, albeit over a relatively narrow wavelength range.

	The radial behaviour of the instrumental Br$\gamma$ index, defined as 
-2.5 log(Br$\gamma$/$2.26\mu$m continuum), is shown in the lower panel of 
Figure 1, where azimuthal averages in 0$\farcs$1 annuli are plotted. The 
Br$\gamma$ and $J-K$ color profiles have qualitatively similar behaviour when 
$r \geq 0\farcs 3$, suggesting that the Br$\gamma$ index and $J-K$ track
similar light sources at these radii, although the Br$\gamma$ curve shows 
sharper structure, due to the higher angular resolution of these data. The 
difference in angular resolution is most apparent near the nucleus, where the 
Br$\gamma$ index dips below the value seen at $2\farcs 0$, whereas $J-K$ near 
the nucleus is redder than what is seen at larger radii.

	Studies of any resolved objects in the circumnuclear 
region of M81 may provide additional clues into the nature of this area. 
Devereux {\it et al.} (1997) used F547M images centered on a WFPC2 wide-field 
array ($0\farcs 1$ pix$^{-1}$) to search for faint sources near the center of 
M81 with negative results. The F547M/PC data considered here have finer pixel 
sampling, and hence are better suited to probing the circumnuclear 
regions of this galaxy. However, the bright nucleus and steep bulge 
brightness profile near the center of M81 complicate efforts to search for 
faint objects. Light from the bulge was removed by first smoothing the PC image 
with a running median filter, and then subtracting the smoothed frames from the 
initial image. The smoothing length was set at twice the FWHM, and 
experiments with different smoothing lengths indicated that the ability to 
detect faint sources was not affected by this parameter. The nucleus was 
subtracted by assuming that it is a radially symmetric point source.

	The central 3$\farcs 4 \times 3\farcs 4$ ($52 \times 52$ pc) portion of 
the F547M/PC image, with the bulge and nucleus removed using the procedure 
described above, is shown in Figure 3. A number of point sources, the brightest 
of which are marked with squares and circles, are present in this figure. These 
sources may not be single luminous stars, but could be resolution elements 
that contain a fortuitous concentration of moderately bright giants. One way 
to estimate the amplitude of such stellar content flucuations is to measure 
the random noise level which, because of the high central surface 
brightness of M81, is dominated by these events. Adopting a conservative 
criterion that only those sources exceeding the random noise level by $5\sigma$ 
are physically real objects, so that only a single noise event may be 
mis-identified as a star in the KIR field, then almost all the sources 
marked in Figure 3 are likely stellar content flucuations and not single bright 
objects; indeed, the sources marked with squares in Figure 3, which have $V 
\sim 23 \pm 0.5$ based on the calibration of Holtzman {\it et al.} (1995), 
exceed the background noise at only the $4\sigma$ level. There is one 
exception, which is the extended object, or group of objects, marked with the 
circle and located 0$\farcs$45 northeast of the main nucleus. This source has 
$V \sim 20.5 \pm 0.5$, and exceeds the noise level by $30\sigma$. Consequently, 
there is a very good chance that it is a physically real object and not a fluke 
superposition of stars. 

	Another way to evaluate the amplitude of stellar 
content flucuations near the M81 nucleus is to co-add fields at larger radii 
to simulate areas of higher surface brightness. Experiments 
of this nature further confirm that the objects 
marked with the squares in Figure 3 are likey stellar 
content fluctuations, and not individual bright sources. These 
experiments were unable to reproduce an object like that 0$\farcs$45 northeast 
of the nucleus, providing further evidence that it is a real source.

	The F160W/NIC2 image, with the bulge and central source removed 
using the procedure described above, is shown in Figure 4. 
The extended source $0\farcs 45$ northeast of the nucleus 
is clearly evident in Figure 4, adding further confidence that this is 
a real object. 

\subsection{A Search for Globular Clusters in the Inner Bulge of M81}

	Crowding makes it difficult to resolve even the most luminous red 
giants near the centers of galaxies outside the Local Group. However, globular 
clusters are sufficiently bright that they should be visible in relatively 
high-density environments. The globular clusters that have been discovered in 
the central regions of the Milky Way are compact, with half light radii in the 
range $1 - 2$ pc (van den Bergh 1994). Objects such as these would subtend 
$0\farcs06 - 0\farcs12$ at the distance of M81, and hence would appear as 
slightly extended sources in the F547M WFPC2 data. With a typical brightness of 
M$_V \sim -7$, bright globular clusters would thus appear as $10\sigma$ sources 
with respect to stochastic fluctuations in stellar content near the center 
of M81; consequently, it is likely that the flucuations with $V \sim 
23$ marked in Figure 3 are not globular clusters.

	Devereux {\it et al.} (1997) concluded that there are no globular 
clusters within 2$^{"}$ of the M81 nucleus, and the archival WFPC2 data 
can be used to extend this search to a much larger area.
The HST F547W/PC image, which covers a $37^{"} \times 
37^{"}$ area, was searched for globular clusters after removing the bulge 
light profile using the procedure described in \S 3.1. The FIND routine in 
DAOPHOT (Stetson 1987), with the `sharp' and `round' parameters set to permit 
elongated and extended objects to be detected, was used to construct a 
preliminary list of objects. Only sources that had at least one pixel 
exceeding the background noise level at the $5\sigma$ level were retained to 
avoid detecting stochastic flucuations in stellar content. The radial light 
profile of each object was then examined by eye to classify the sources as 
either stellar or extended. The brightnesses of individual objects were 
measured with the DAOPHOT aperture photometry routine PHOT. Only one extended 
object, located $15\farcs6$ north west of the nucleus, was found outside the 
circumnuclear region in the PC field. This source has V=20.2, which 
corresponds to M$_V = -8.2$ if A$_V = 0.9$ mag (Perelmuter \& Racine 1995).
Younger star clusters and HII regions will also appear as 
extended objects and, lacking the color information needed to identify such 
sources, we can not unambiguously state whether or not this object is a 
bright globular cluster.

	A larger population of globular cluster candidates have been discovered 
with the three WF detectors. Neglecting obvious bright foreground stars, 
the technique described above finds 25 sources with $V \leq 21.5$, 
which is the estimated completeness limit of these data, and 15 of these are 
extended objects. The WFPC2 field covers $\sim 1/3$ of the projected area 
within $139\farcs$0 ($\sim 2$kpc) from the center of M81. Consequently, we 
estimate that there are $3 \times 15 \pm 4 = 45 \pm 12$ extended objects with 
M$_V \leq -7$ within 2 kpc of the galaxy center.

\section{DISCUSSION AND SUMMARY}

	Near-infrared images obtained with the CFHT adaptive optics system have 
been combined with archival HST data to investigate the central regions of the 
nearby spiral galaxy M81. Studies of the central forbidden emission line 
spectrum of this galaxy predict that kinematically distinct structures 
(Filippenko \& Sargent 1988, Keel 1989) and significant density stratification 
(Ho {\it et al.} 1996) should occur in the NLR, which may extend out to a few 
arcsec from the nucleus (Filippenko \& Sargent 1988). We find that the 
near-infrared SED near the M81 nucleus changes with radius, providing 
additional evidence of stratification in the circumnuclear interstellar 
medium. $J-K$ peaks $\sim 0\farcs4$ from the nucleus, at a value that is 
significantly redder than what is seen in the corresponding portion of M31, 
although at $1\farcs 5$ from the nucleus M31 and M81 have similar $J-K$ colors. 
The CO index near the nucleus is relatively weak, indicating that the 
red $J-K$ color is not due to a population of extremely red stars. 

	The near-infrared SED of the M81 circumnuclear region can be reproduced 
with a simple two component model, in which emission from warm dust, which is a 
common occurence in LINER (Wilner {\it et al.} 1985) and Seyfert 
(Alonso-Herrero {\it et al.} 1996, and references therein) systems, is added to 
a stellar component. Following Alonso-Herrero {\it et al.} (1996), we have 
generated model SEDs for various mixtures of dust and stars, and the results 
are shown as dashed lines in Figure 2. The near-infrared colors of 
the stellar component were fixed at the values in M81 $1\farcs$5 from the 
nucleus, while the dust is assumed to follow a black-body curve with a 
temperature of 1000K. The dust temperature is based on 
observations of other Seyferts at longer wavelengths (Alonso-Herrero {\it et 
al.} 1996), although at near-infrared wavelengths the 
model SEDs are insensitive to this parameter.

	The simple two component models reproduce the trends defined by the 
observations on the $(J-H, H-K)$ and $(CO, J-K)$ TCDs, and predict that 20\% of 
the light in $K$ within $0\farcs5$ of the M81 nucleus originates from warm 
dust. At small distances from the nucleus the observations deviate from the 
predicted trend, due to continuum emission from the nuclear source. The 
presence of hot dust near the center of M81 is 
yet another example of how this galaxy harbours a small-scale version of the 
engines that occur in more powerful systems. 

	The central regions of AGNs are morphologically complex environments 
(e.g. Malkan, Gorjian, \& Tam 1998), and we have used 
archival F547M and F160W HST images to find a 
faint extended object in the M81 NLR with $V \sim 20.5$, which corresponds to 
M$_V \sim -7.3$ if A$_V = 0.3$ (Filippenko \& Sargent 1988) and $\mu_0 = 27.5$. 
The complicated PSF in the CFHT images makes it difficult to investigate this 
structure with these data, and the nature of this object remains a matter of 
speculation. However, the fact that it is detected in both the WFPC2 and 
NIC2 datasets suggests that it has a flat SED, so a source 
dominated solely by either very red or blue stars can be ruled out. 
There is evidence for recent star formation in the central regions of 
Seyfert-1 galaxies (e.g. Gonzalez Delgado \& Perez 1997), and we 
speculate that this object could be an HII region. High angular 
resolution spectra at optical wavelengths, which would allow key 
diagnostics such as H$\alpha$ and [OIII] emission to be studied, will provide 
insight into the nature of this object. The detection of a star-forming 
region near the center of M81 may have implications for the origins of the 
young stellar complexes that are found within a few tens of pc of the 
GC (e.g. Morris \& Serabyn 1996, and references therein).

	We have also searched for globular clusters in the central regions 
of M81. This galaxy is an interesting target for studying the central 
globular cluster content of spiral galaxies as it is relatively 
close while the inner regions are more-or-less free 
of the absorption that plagues efforts to survey the corresponding area 
of the Milky Way. Perelmuter \& Racine (1995) investigated 
the global properties of the M81 globular cluster system, and found that the 
spatial distribution of clusters at large radii is similar to 
that of the Galactic and M31 globular cluster systems. However, the 
Perelmuter and Racine (1995) survey excluded the region within 1 arcmin 
of the nucleus, so it is not known if there is an absence of 
globular clusters near the galaxy center, as is the 
case in the Milky-Way and M31 (Battistini {\it et al.} 1993).

	Using archival WFPC2 data, we estimate that there are $\sim 45 \pm 12$ 
extended objects with M$_V \leq -7$, many of which will be globular clusters, 
within 2 kpc of the galaxy center. For comparison, there are 17 clusters with 
M$_V \leq -7$ within 2kpc of the Galactic Center (Harris 1996). The M81 
cluster system is $1.4 \pm 0.2$ times more populous than that in the Milky-Way 
(Perelmuter \& Racine 1995), so $24 \pm 3.4$ globular 
clusters would be expected within 2 kpc of the center of M81 if the 
central spatial distributions of the M81 and Galactic cluster systems scaled 
according to total cluster population. Given that the difference between the 
observed and predicted number of clusters, $21 \pm 13$, is significant at less 
than the $2\sigma$ level, and that some of the extended objects detected 
in M81 will not be globular clusters, these data support the hypothesis 
that the central spatial distributions of the two cluster systems are similar.

	There is only one globular cluster candidate within 0.27 kpc of the 
center of M81, suggesting that the number density of clusters does not build 
towards very small radii in this galaxy. The near absence of bright 
globular clusters within a few 100 pc of the centers 
of M31 and M81, the nearest large external spiral galaxies, strongely suggests 
that the dearth of clusters in the corresponding part 
of the Galaxy is a real effect, and not an artifact of 
extinction-induced incompleteness or a statistical fluke.

	It is not clear if a single physical mechanism is responsible for the 
absence of globular clusters near the centers of spiral galaxies. Murali \& 
Weinberg (1997) and Vesperini (1997) argue that dynamical processes play a 
major role in shaping the observed properties of the Galactic cluster system, 
and that clusters which formed in the innermost regions of the Galaxy have been 
disrupted. However, the central globular cluster content in elliptical galaxies 
is deficient with respect to the underlying galactic light, and the 
Milky-Way cluster system falls along the relation between system core radius 
and galaxy luminosity defined by early-type galaxies (Forbes {\it et al.} 
1996), suggesting that the processes that shaped the properties of the globular 
cluster systems in the Milky-Way and elliptical galaxies are related. This 
connection is significant because Lauer \& Kormendy (1986) argue that the 
spatial distribution of globular clusters in very large ellipticals, such 
as M87, can not be the result of dynamical friction, and hence may have been 
imprinted at early epochs. Nevertheless, Lauer \& Kormendy (1986) also note 
that dynamical friction will still be a factor for clusters located well within 
the globular cluster system core radius. The absence of clusters near the 
centers of nearby spiral galaxies may thus be the result of the combined 
effects of an initially low central cluster density that was subsequently 
culled by dynamical effects.

\vspace{0.3cm}
	Sincere thanks are extended to Sidney van den Berg, Jim Hesser, Simon 
Morris, and Sylvain Veilleux for commenting on an earlier draft of this paper. 
An anonymous referee also made suggestions that greatly improved the 
manuscript. Thanks are also extended to Daniel Durand and Severin Gaudet for 
providing the pipeline-processed NICMOS data used in this paper.


\clearpage

\begin{table*}
\begin{center}
\begin{tabular}{lcc}
\tableline\tableline
Filter & Exposure Time & FWHM \\
 & (sec) & (arcsec) \\
\tableline
J & $16 \times 90$ & 0.34 \\
H & $16 \times 90$ & 0.31 \\
K & $16 \times 90$ & 0.20 \\
Br$\gamma$ & $8 \times 60$ & 0.17 \\
$2.26\mu$m continuum & $8 \times 60$ & 0.17 \\
CO & $8 \times 90$ & 0.17 \\
\tableline
\end{tabular}
\end{center}
\caption{DETAILS OF OBSERVATIONS}
\end{table*}

\clearpage
\begin{center}
FIGURE CAPTIONS
\end{center}

\figcaption
[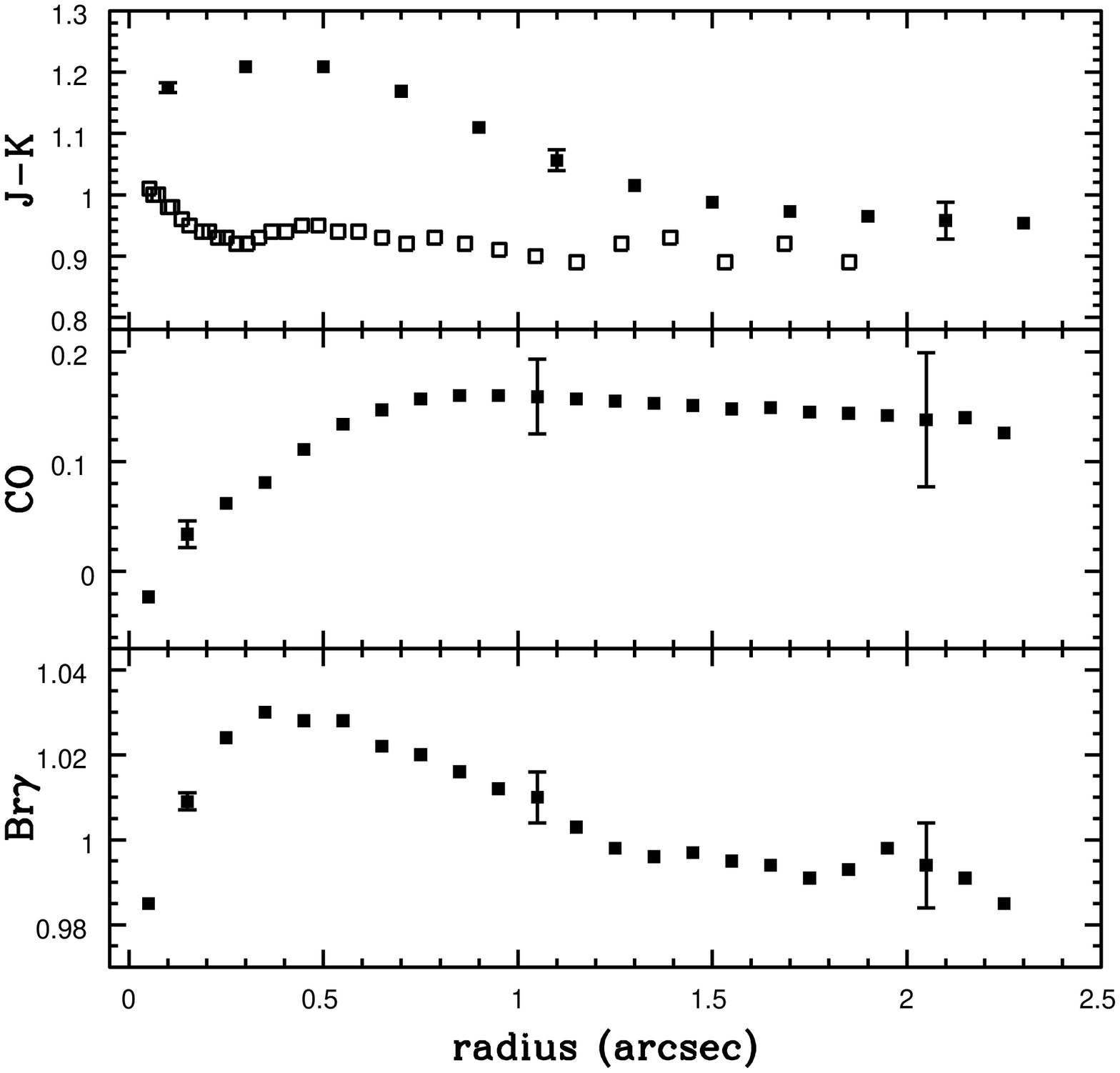]
{Radial $J-K$, CO, and Br$\gamma$ profiles near the center of M81, with 
distance measured from the galaxy nucleus. The M81 data, shown 
as solid squares, are azimuthal averages, and the error bars show the 
systematic errors introduced by uncertainties in the background. 
The open squares in the top panel are major axis 
color measurements for M31 taken from Table 1 of Mould {\it et al.} (1989), 
shifted to the distance of M81 by assuming that $\mu_0^{M31} = 24.5$ and 
$\mu_0^{M81} = 27.5$. The CO and Br$\gamma$ indices are in magnitude units, 
with the latter in the instrumental system.}

\figcaption
[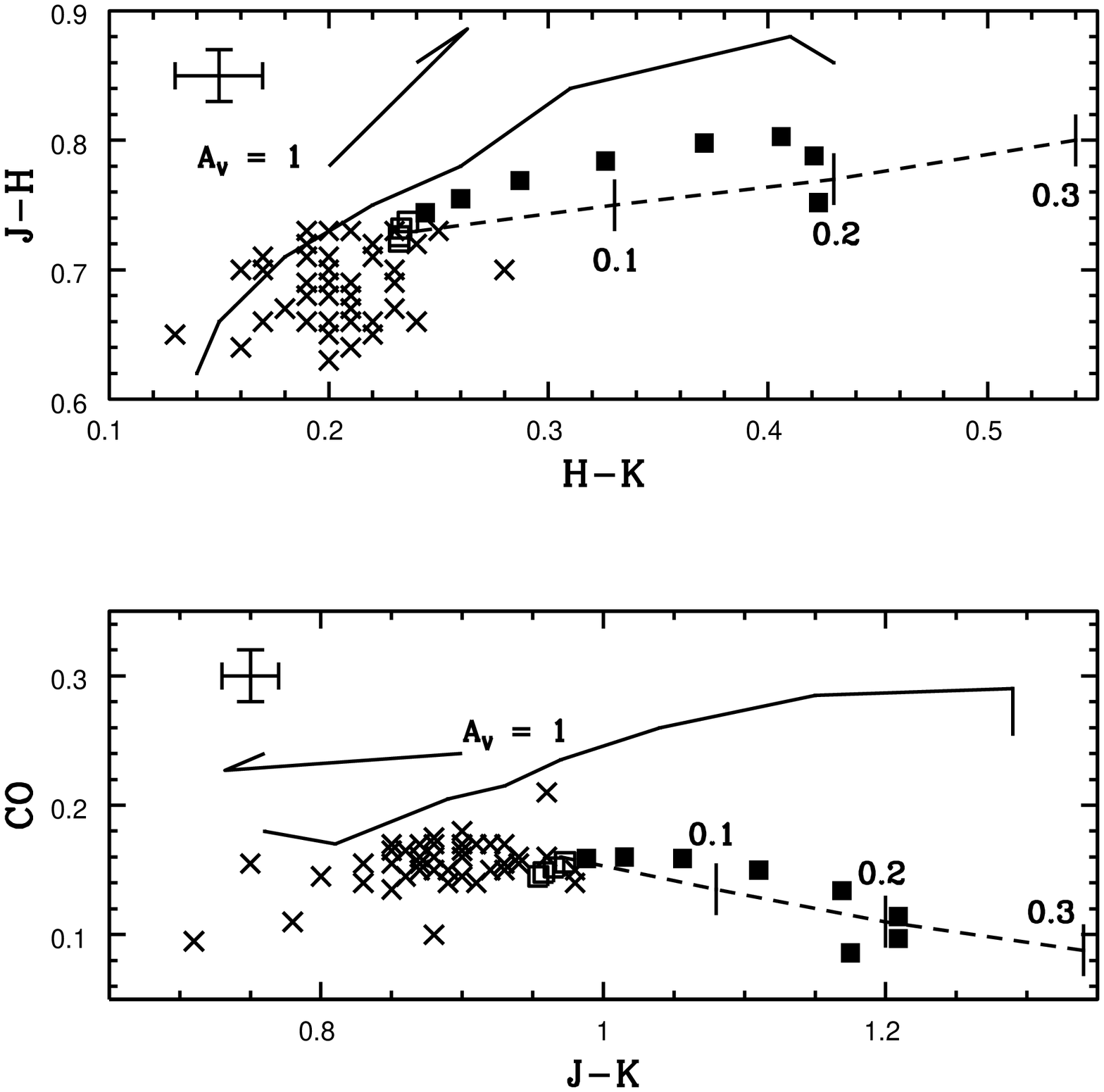]
{$(J-H, H-K)$ and $(CO, J-K)$ diagrams for the central regions of M81. The data 
plotted are azimuthal averages in 0$\farcs$2 annuli centered on the M81 nucleus 
with filled and open squares refering to distances $\leq 1\farcs$5 and $\geq 
1\farcs$5 from the galaxy center, respectively. There is almost a one-to-one 
correspondence between color and distance from 
the M81 nucleus, in the sense that points with the 
largest $H-K$ and $J-K$ values are closest to the galaxy center. Also shown are 
measurements for elliptical and S0 galaxies from Table 4 of Frogel {\it et al.} 
(1978) (crosses), and the trend defined by M giants in BW from Table 3B of 
Frogel \& Whitford (1987) (solid line). The error bar in each panel shows the 
$1\sigma$ uncertainties in the photometric zeropoints derived from the standard 
star measurements. Reddening vectors with lengths approriate for A$_V = 1$ mag, 
based on the Rieke \& Lebofsky (1985) reddening curve and $E(CO)/E(B-V) = 
-0.04$ (Elias, Frogel, \& Humphreys 1985), are shown. The dashed lines show the 
effects of combining the near-infrared SED of M81, as infered from data at 
radii in excess of $1\farcs$5 from the nucleus, with a T$_{eff} = 1000$ K 
black-body. Points showing ratios of black-body to stellar flux of 0.1, 0.2, 
and 0.3 in $K$ are marked.}

\figcaption
[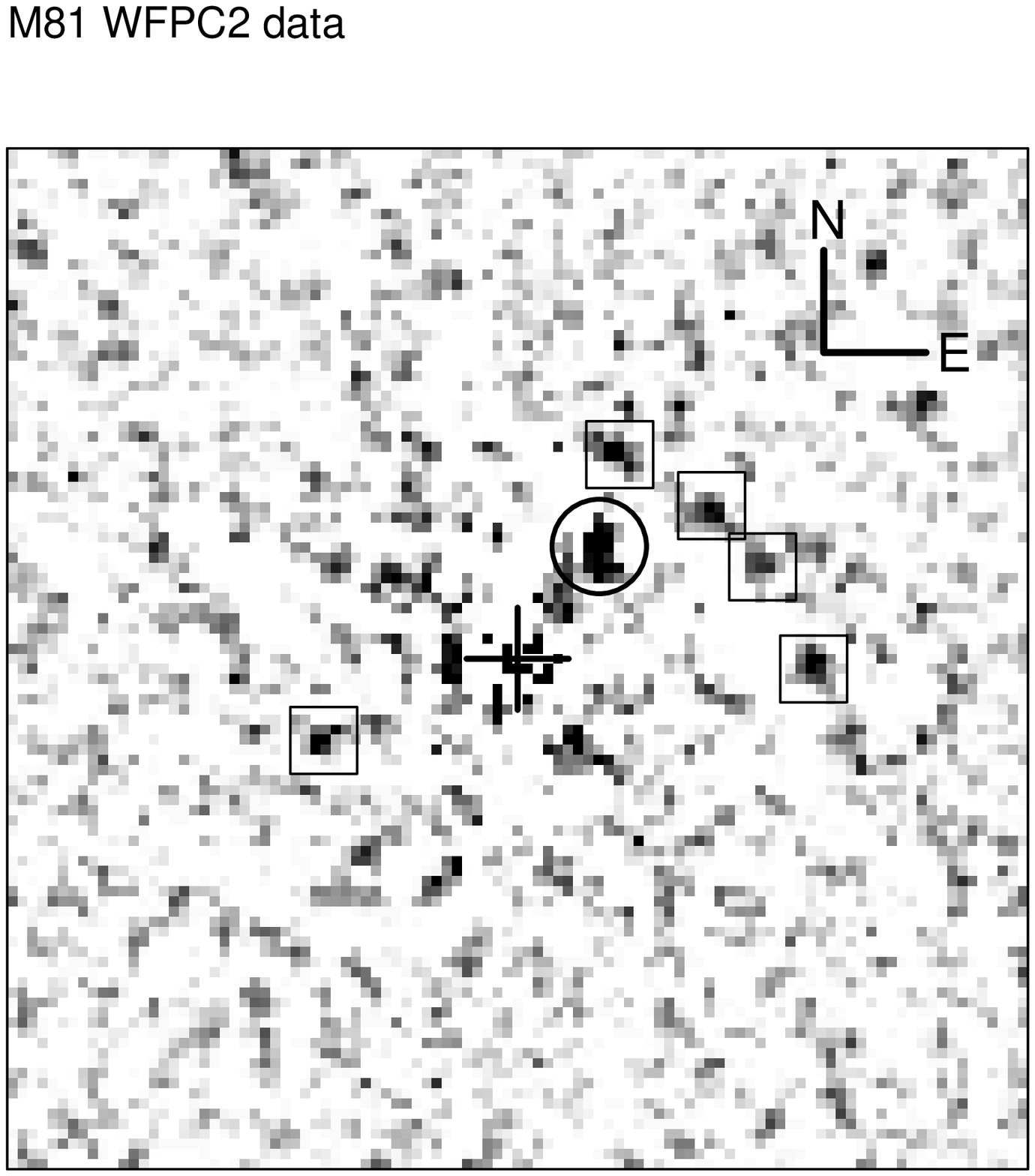]
{The portion of the F547M Planetary Camera image showing the central 3$\farcs 4 
\times 3\farcs 4$ ($52 \times 52$ pc) of M81. The bulge light profile and the 
central nuclear source have been removed using the procedure described in the 
text, and the cross marks the center of M81. 
North is at the top, and east is to the right. 
The brightest objects have the darkest colors. The extended 
object marked with a circle is located $0\farcs 45$ north east of the 
nucleus, and exceeds the random noise in the background at the $30\sigma$ 
level. The squares indicate the next five brightest objects in the field, 
which are likely stochastic fluctuations in stellar content, rather than 
individual bright stars.}

\figcaption
[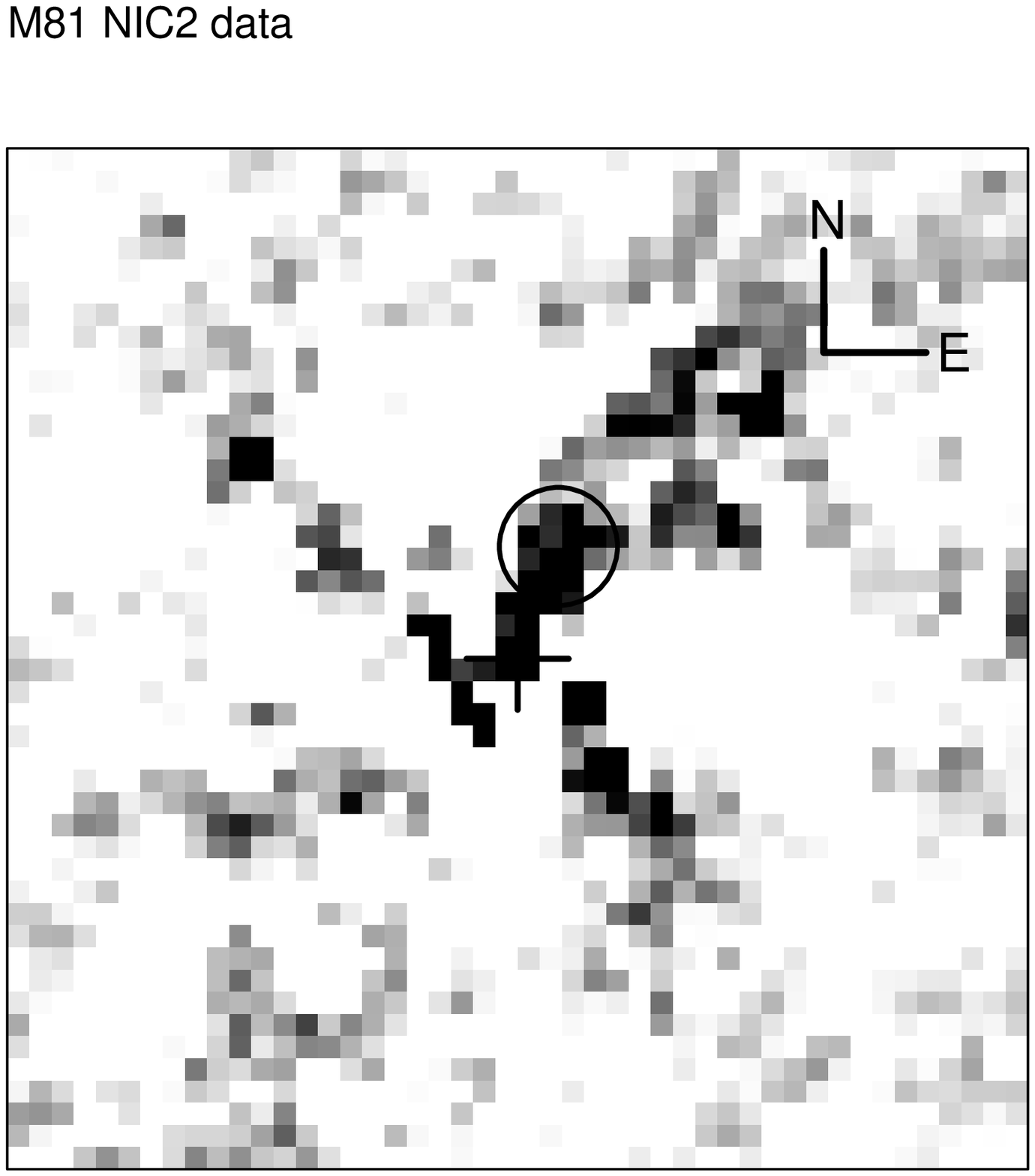]
{The portion of the F160W NIC2 image that corresponds to the field shown in 
Figure 3. The bulge light profile and the central nuclear source have been 
removed using the procedure described in the text, and the cross 
marks the center of M81. North is at the top, and east is to the right. 
The brightest objects have the darkest colors. The circle marks the extended 
object indicated in Figure 3, which is located $0\farcs 45$ north east 
of the nucleus.}

\clearpage
\begin{references}

\reference{} Alonso-Herrero, A., Ward, M. J. \& Kotilainen, J. K. 1996, MNRAS, 
278, 902

\reference{} Bartel, N. {\it et al.} 1982, ApJ, 262, 556

\reference{} Battistini, P. L. {\it et al.} 1993, A\&Ap, 272, 77

\reference{} Bessell, M. S., \& Brett, J. M. 1988, PASP, 100, 1134

\reference{} Bietenholz, M. F. {\it et al.} 1996, ApJ, 604

\reference{} Brocklehurst, M. 1971, MNRAS, 153, 471

\reference{} Casali, M., \& Hawarden, T. 1992, UKIRT-JCMT Newsletter, 4, 33

\reference{} Chapman, S. C., Walker, G. A. H., \& Morris, S. L. 1998, 
astro-ph 9810250

\reference{} Davidge, T. J. 1997, AJ, 113, 985

\reference{} Davidge, T. J. 1998, AJ, 115, 2374

\reference{} Davidge, T. J., Simons, D. A., Rigaut, F., Doyon, R., \& 
Crampton, D. 1997a, AJ, 114, 2586

\reference{} Davidge, T. J., Rigaut, F., Doyon, R., \& Crampton, D. 1997b, 
AJ, 113, 2586

\reference{} Devereux, N., Ford, H., \& Jacoby, G. 1997, ApJ, 481, L71

\reference{} Eckart, A., Genzel, R., Hofmann, R., Sams, B. J., \& 
Tacconi-Garman, L. E. 1995, ApJ, 445, L23

\reference{} Elias, J. H., Frogel, J. A., \& Humphreys, R. M. 1985, APJS, 57, 
91
 
\reference{} Faber et al. 1997, AJ, 114, 1771

\reference{} Falcke, H. 1996, ApJ, 464, L67

\reference{} Filippenko, A. V., \& Sargent, W. L. W. 1988, ApJ, 324, 134

\reference{} Forbes, D. A., Franx, M., Illingworth, G. D., \& Carollo, 
C. M. 1996, ApJ, 467, 126

\reference{} Frogel, J. A., \& Whitford, A. E. 1987, ApJ, 320, 199

\reference{} Frogel, J. A., Persson, S. E., Aaronson, M., \& Matthews, K. 
1978, ApJ, 220, 75

\reference{} Giles, K. 1977, MNRAS, 180, 57P

\reference{} Gonzalez Delgado, R. M., \& Perez, E. 1997, MNRAS, 284, 931

\reference{} Haller, J. W., Rieke, M. J., Rieke, G. H., Tamblyn, P., Close, L. 
\& Melia, F. 1996, ApJ, 456, 194

\reference{} Harris, W. E. 1996, AJ, 112, 1487

\reference{} Ho, L. C., Filippenko, A. V., \& Sargent, W. L. W. 1996, ApJ, 
462, 183

\reference{} Holtzman, J. A., {\it et al.} 1995, PASP, 107, 1065

\reference{} Keel, W. C. 1989, AJ, 98, 195

\reference{} Krabbe, A. et al. 1995, ApJ, 447, L95

\reference{} Larkin, J. E., Armus, L., Knop, R. A., Soifer, B. T., \& 
Matthews, K. 1998, ApJS, 114, 59

\reference{} Lauer {\it et al.} 1993, AJ, 106, 1436

\reference{} Lauer, T. R., \& Kormendy, J. 1986, ApJ, 303, L1

\reference{} Malkan, M. A., Gorjian, V., \& Tam, R. 1998, ApJS, 117, 25

\reference{} Morris, M., \& Serabyn, E. 1996, ARAA, 34, 645

\reference{} Mould, J., Graham, J., Matthews, K., Soifer, B. T., \& Phinney, 
E. S. 1989, ApJ, 339, L21

\reference{} Murali, C., \& Weinberg, M. D. 1997, MNRAS, 288, 749

\reference{} Perelmuter, J-M, \& Racine, R. 1995, AJ, 109, 1055

\reference{} Reuter, H.-P., \& Lesch, H. 1996, A\&A, 310, L5

\reference{} Rieke, G. H., \& Lebofsky, M. J. 1985, ApJ, 288, 618

\reference{} Rigaut, F. {\it et al.} 1998, PASP, 110, 152

\reference{} Sandage, A., \& Tammann, G. A. 1987, A Revised Shapley-Ames 
Catalog of Galaxies, Carnegie, Washington DC.

\reference{} Sellgren, K., McGinn, M. T., Becklin, E. E., \& Hall, D. N. B. 
1990, ApJ, 359, 112

\reference{} van den Bergh, S. 1994, AJ, 108, 2145

\reference{} Vesperini, E. 1997, MNRAS, 287, 915

\reference{} Willner, S. P., Elvis, M., Fabbiano, G., Lawrence, A., \& Ward, 
M. J. ApJ, 299, 443

\end{references}
\end{document}